\def\comment#1{}

\newcommand{\beg}{\begin{eqnarray}}
\newcommand{\eee}{\end{eqnarray}}

\documentstyle[prl,aps,amsfonts,psfig,multicol,epsf]{revtex}
\def\cm#1{}

\begin{document}
\title{ %Massless boson, ``neutral"  vortices and
Vortices with fractional flux   \\% carrying an arbitrary fraction of magnetic
%flux quantum \\ 
in two-gap superconductors and in extended Faddeev model}
%\vskip 0.3cm
\author{
Egor Babaev 
\thanks{email: egor@nordita.dk \  http://www.teorfys.uu.se/PEOPLE/egor/ }
}
\address{
     Institute for Theoretical Physics, Uppsala University, Box 803, 75108 Uppsala, Sweden \\
    NORDITA, Blegdamsvej 17, DK-2100 Copenhagen, Denmark \\
Science Institute, University of Iceland, Dunhaga 3, 107 Reykjavik, Iceland}
\maketitle
\begin{abstract}
We discuss linear topological defects allowed in  
two-gap superconductors and equivalent extended Faddeev model.
We show that in these systems there exist vortices 
which carry an arbitrary fraction of magnetic flux quantum.
Besides that we discuss topological defects 
which do not carry magnetic flux  and
 describe features of ordinary one-magnetic-flux-quantum vortices
in the two-gap  system.
%We  also discuss  an experiment where  these vortices carrying a  fraction 
%of magnetic flux quantum can be created and observed. 
The results could be relevant for the newly discovered two-band superconductor $Mg B_2$.
 \end{abstract}
\newcommand{\la}{\label}
\newcommand{\aaa}{\frac{2 e}{\hbar c}}
\newcommand{\Pfaff}{{\rm\, Pfaff}}
\newcommand{\kA}{{\tilde A}}
\newcommand{\G}{{\cal G}}
\newcommand{\cP}{{\cal P}}
\newcommand{\M}{{\cal M}}
\newcommand{\E}{{\cal E}}
\newcommand{\btd}{{\bigtriangledown}}
\newcommand{\W}{{\cal W}}
\newcommand{\X}{{\cal X}}
\renewcommand{\O}{{\cal O}}
\renewcommand{\d}{{\rm\, d}}
\newcommand{\bfi}{{\bf i}}
\newcommand{\e}{{\rm\, e}}
\newcommand{\bfx}{{\bf \vec x}}
\newcommand{\bfn}{{\bf \vec n}}
\newcommand{\bfE}{{\bf \vec E}}
\newcommand{\bfB}{{\bf \vec B}}
\newcommand{\bfv}{{\bf \vec v}}
\newcommand{\bfU}{{\bf \vec U}}
\newcommand{\ccc}{{\vec{\sf C}}}
\newcommand{\bfp}{{\bf \vec p}}
\newcommand{\f}{\frac}
\newcommand{\bfA}{{\bf \vec A}}
\newcommand{\non}{\nonumber}
\newcommand{\be}{\begin{equation}}
\newcommand{\ee}{\end{equation}}
\newcommand{\ba}{\begin{eqnarray}}
\newcommand{\ea}{\end{eqnarray}}
\newcommand{\bastar}{\begin{eqnarray*}}
\newcommand{\eastar}{\end{eqnarray*}}
\newcommand{\half}{{1 \over 2}}
\newcommand{\qq}{{\frac{|\Psi_1|^2}{m_1}}}
\newcommand{\ww}{{\frac{|\Psi_2|^2}{m_2}}}
\begin{multicols}{2}
\narrowtext
A fundamental property of the Abelian Higgs model
is the quantization of magnetic flux  \cite{aaa}.
In an ordinary superconductor the Abrikosov vortices can carry 
only integer number of magnetic flux quanta.
The intriguing possibilities of topological defects carrying 
a fraction of flux quantum have long attracted interest 
 and several nontrivial realizations 
were identified. For example,  a half flux-quantum vortex in
a spin-1 condensate is a configuration where a Cooper
pair has its spin reversed when moving around the vortex
core (this is an analogue of  an Alice string in high energy physics 
where a particle moving around the string flips its charge
or enters a ``shadow world")  also a half flux-quantum 
vortex can be formed on a junction of 3 grain 
boundaries in a crystal etc \cite{volovik}.
In this paper we discuss vortices in
two-gap superconductors  \cite{mult,legg} (known in particle
physics as a Higgs doublet model \cite{tdlee}) and in the extended Faddeev
model. We show that these vortices
can carry an arbitrary fraction of magnetic flux quantum.

%The two-gap superconductivity was first discussed a long time ago  .
%Analogous model has been studied in particle 
%physics where it is known as ``Higgs doublet" (see e.g. \cite{tdlee}).
Experimentally,  two-gap superconductivity has
been observed in the  transition metals  $Nb$, $Ta$,$V$
and in $Nb$-doped $Sr Ti O_3$ \cite{transi}.
More recent experiments indicate the two-gap nature of superconductivity 
in    $Mg B_2$ \cite{mgb} and $2H-NbSe_2$ \cite{2h}.  
Two-gap models appear also in the  theoretical studies
of   liquid metallic hydrogen, 
which should allow superconductivity 
of both electronic and protonic Cooper pairs \cite{ashc2}.  
In liquid metallic deuterium a deuteron superfluidity 
may be present along with superconductivity of electronic 
Cooper pairs \cite{ashc2}. 
Other realizations of the two-gap system are superconductors
with two types of pairing (e.g. a mixture of $s$- and $p$-wave condensates).

A two-gap superconductor  is   described by a two-flavour 
Ginzburg-Landau   free energy functional:
\beg
&&
F = %\biggl[ 
\frac{1}{2m_1} \left| \left( \nabla +
i e {\bf A}\right) \Psi_1 \right|^2 + 
\frac{1}{2m_2}  \left| \left( \nabla +
i e {\bf A}\right) \Psi_2 \right|^2 
 \nonumber \\
&&
+ { V} (|\Psi_{1,2}|^2)+ \eta [\Psi_1^*\Psi_2+\Psi_2^*\Psi_1] 
+ \frac{{\bf B}^2}{2}
%\biggr]
\la{act}
\eee
where $\Psi_\alpha = |\Psi_\alpha|e^{i \phi_\alpha}$ and $
{V} (|\Psi_{1,2}|^2)=-b_\alpha|\Psi_\alpha|^2+ 
\frac{c_\alpha}{2}|\Psi_\alpha|^4
$ and $\eta $ is a characteristic of the interband Josephson coupling strength
\cite{legg}.
%We consider two oppositely charged scalar fields. The 
%results holds true also for two condensates  
%of similar charge (charge inversion $e \rightarrow -e$ of the 
%field $\Psi_\alpha$ amounts to $|\Psi_\alpha| e^{i \varphi_\alpha}\rightarrow 
%|\Psi_\alpha| e^{-i \varphi_\alpha}$).
%Indeed, there can be  present additional terms in (\ref{act}),
%however the purpose of this Letter is to examine the 
%very simplest 
%form of the model for a two-gap superconductor, because
%as we discuss below even in the form  (\ref{act})
%it  exhibits remarkable somewhat counter-intuitive
%phenomena.

Many exotic properties of  (\ref{act})  are  obscured 
in the Ginzburg-Landau presentation of the free energy functional.
In \cite{we} it was  shown that there exists an exact 
equivalence mapping between the model (\ref{act}) 
and an extended version of Faddeev's $O(3)$ nonlinear $\sigma$-model \cite{fadde},
 which 
describes the two-gap superconductors in terms of {\it gauge invariant }variables
which explicitly show the degrees of freedom present in the system.
This model consists of 
a three-component unit vector $\bfn$ in interaction with a massive 
vector field $\vec{\sf C}$ and a  
density-related variable  $\rho$ \cite{tri}:
\beg
F&=& \frac{\rho^2}{4}(\nabla \bfn)^2 + 
(\nabla \rho)^2 + \f{\rho^2}{16}
 \vec{\sf C}^2
+ {{V}}(\rho , n_3 )+\rho^2 K n_1
\nonumber \\
&+& \frac{1}{32  e^2}
\left(
%\f{1}{\hbar}
\partial_i {\sf C}_j -\partial_j 
{\sf C}_i -\bfn \cdot \partial_i
\bfn \times \partial_j\bfn
\right)^2 
%\nonumber  \\ && 
\la{e3}
\eee
where $\partial_i = \f{d}{dx_i}$, $ { V}= A + B n_3 + C n_3^2$.
The models (\ref{act}) and  (\ref{e3}) are connected in the following 
way \cite{we}: coefficients $A,B,C$ are given by: 
 $ \ A=\rho^2 [  4c_1m_1^2 + 4c_2m_2^2 -b_1m_1 - b_2 m_2 ]; \ 
B= \rho^2 [ 8c_2m_2^2 - 8c_1m_1^2 -b_2m_2 + b_1 m_1]; \
C= 4\rho^2 [c_1m_1^2 + c_2m_2^2]$. The
position of the unit vector $\bfn$ on the sphere $S^2$ can be characterized
by two angles as follows: $ \bfn= (\sin\theta\cos\gamma,\sin\theta\sin\gamma,\cos\theta)
$, 
%The variables of (\ref{act}) and (\ref{e3}) are related in the following way \cite{we}:
where $\gamma=(\phi_1 -\phi_2); |\Psi_{1,2}| = 
\left[\sqrt{2m_1}\ \rho \sin\left(\f{\theta}{2}\right),
\sqrt{2m_2} \ \rho \cos\left(\f{\theta}{2}\right) \right] $;
%\nonumber
%\la{psi}
$\vec{\sf C }=  %\biggl[
\frac{i  }{ m_1 \rho^2}
\left\{\Psi_1^*\nabla \Psi_1-
\Psi_1 \nabla \Psi_1^*\right\}
% \nonumber \\
%&& -\nonumber \\ 
+\frac{i }{m_2  \rho^2}
\{\Psi_2^*\nabla \Psi_2- 
\Psi_2 \nabla \Psi_2^*\} -
\frac{2e}{\rho^2} \left( \f{|\Psi_1|^2}{m_1} 
+\f{|\Psi_2|^2}{m_2} \right){\bf A}%\biggr]$
$.
% - \frac{2 e^2}{m}|\Psi_i|^2 {\bf A}
%\nonumber
%\la{A}
%\eee
%The vector field $\ccc$ is directly related to supercurrent density $\bf J$:  $ \ccc = { \bf J}/{e \rho^2} $.
We consider the system 
in the presence of an interband Josephson coupling 
$\eta[\Psi_1^*\Psi_2+\Psi_2^*\Psi_1] = \rho^2 K n_1 $
where $K\equiv 2\eta\sqrt{m_1m_2}$.
The potential term $V$ in (\ref{e3}) determines the energetically preferred 
ground state value for $n_3$, which corresponds to 
uniform density of both condensates. We denote it as
% (we denote it   by ${\tilde n_3}$) \cite{we}:
${ \tilde n_3} \equiv \left[\frac{ N_2}{m_2}-\frac{N_1}{m_1}\right]\left[\frac{  N_1}{m_1} + \frac{N_2}{m_2}\right]^{-1}$
%\label{nstar}
%\ee
where $N_{1,2}$ stands for the average $<|\Psi_{1,2}|^2>$.
Thus, {\it in the presence of the intrinsic Josephson effect},
 the ground state value of $\bfn$ corresponds to a point where $n_1=-1$
on  a circle specified   by the condition $n_3={\tilde n}_3 \equiv \cos\tilde{\theta}$.
We  note that (\ref{e3})  shows the Meissner effect,
which is manifest in generation of the mass  for $ \vec{\sf C}$ \cite{we}.
% One may obsevre that  London equation for magnetic field follows 
%from (\ref{e3})  in the  London limit ($|\Delta_{1,2}|={\em const}$):  
%First one can obsevre that  the  term $(\rho^2/16)  \vec{\sf C}^2$ has a natural physical 
% interpratation being  the kinetic energy density of supercurrent $\bf J$ of Cooper pairs 
% $W_{kin}=1/2(m_1/N_1+m_2/N_2)({\bf J}^2/e^2)$   [from (\ref{A}) one can easily see that 
%$\ccc=e\rho^2 {\bf J}$]. Taking into account Maxwell equation ${\rm curl}{\bf H} = 4 \pi/c {\bf J}$,
%the contribution to the free energy functional  from  the energy of external magnetic field and kinetic 
%energy of screening Meissner current in the London limit is:
% \be E_H =\f{1}{8\pi} \int d x[{\bf H}^2 + \lambda^2{{\rm curl} {\bf H}}^2]\ee
The corresponding length scale  $\lambda$ is  the London magnetic field 
penetration length:  $ \lambda^2=\frac{1}{ e^2}\left[ \f{|\Psi_1|^2}{m_1}+
\f{|\Psi_2|^2}{m_2}\right]^{-1}$. %=1/(4 e^2\rho^2)$.
% A standard variation procedure gives exactly  the London equation  for the magnetic field
%\be {\bf H} + \lambda^2{\ {\rm curl} \ {\rm curl} \ {\bf H}}=0  \la{london}. \ee 
%We should remark that actually the London limit  (that is, the limit when 
%$|\Psi_{1,2}| = const$)
%misses  fundamentally important   physical feature of
%a two-gap system, namely the contribution 
%to the magnetic field of the term $\bfn \cdot \partial_i
%\bfn \times \partial_j\bfn$ in (\ref{e3})  \cite{we}. Existence 
%of this term for example allows the system to form stable 
%knot solitons \cite{we}. The effects
%discussed in the present paper however can be described 
%adequately in  the London limit.
%The extended Faddeev model (\ref{e3}) 
%displays structure of a two-gap  superconductor
%in particularly convenient 
%and simple intuitive way.
One can  observe  from (\ref{e3})   that along 
with the Meissner effect in a two-band superconductor 
there exists  a  neutral boson.
That is, if one considers a uniform density of condensates, 
the vector field $\vec{\sf C}$  decouples from the field $\bfn$, 
because $\bfn \cdot \partial_i \bfn \times \partial_j\bfn \propto \sin\theta[\partial_i \gamma\partial_j \theta-\partial_i \gamma\partial_j \theta]  = 0 $,
 when $n_3=\cos\tilde{\theta} = const$. 
%The  vector $\bfn$ in this case resides of the circle on the unit sphere $S^2$
%defined by the condition $n_3 = \tilde{n}_3$. 
Thus, when 
$\eta=0$, the system possesses  a
massless neutral $O(2)$ excitation associated with the phase variable 
$ \gamma =(\phi_1 -\phi_2)$.
This phenomenon has no counterpart in  one-gap superconductors where the Goldstone
boson associated with $O(2)$ symmetry is ``eaten" by the Meissner effect.
{ \it In the presence of the Josephson effect ($\eta \ne 0$)  the variables  still
decouple but the Josephson  term $\rho^2 K n_1$ breaks the neutral
$O(2)$ symmetry by giving a nonzero mass to $n_1$}.

Let us discuss  vortices in the system (\ref{e3}).
In the London limit ($|\Psi_i| = const $),  the eqs. 
(\ref{act}), (\ref{e3}) become:
\beg
&&F= \frac{\rho^2}{4}(\nabla \bfn)^2 + 
\f{\rho^2}{16}\vec{\sf C}^2 +
\frac{1}{32  e^2}
[\partial_i {\sf C}_j -\partial_j 
{\sf C}_i]^2 + \rho^2 K n_1=
 \nonumber \\
&&\f{\rho^2 }{4} \sin^2\tilde{\theta} (\nabla \gamma)^2 +
\rho^2 \Bigl[ \sin^2\Bigl( \f{\tilde{\theta}}{2}\Bigr) \nabla \phi_1
 + \cos^2\Bigl( \f{\tilde{\theta}}{2}\Bigr)\nabla \phi_2 
% \nonumber \\
- e{\bf A}\Bigr]^2
\nonumber \\
&&
+ \f{{\bf B}^2}{2}
+ \rho^2 K \sin\tilde{\theta} \cos\gamma 
\la{new1}
\eee

From this expression one can observe that the  vortices characterized by 
$\Delta( \phi_1 +\phi_2)  \equiv  \oint_\sigma d l [\nabla( \phi_1 +\phi_2)]=4 \pi m;  \ \  \Delta( \phi_1 - \phi_2)  =0$,
(where we integrate over a closed curve $\sigma$
around the vortex core)
is the  analogue of $m$-flux quanta Abrikosov 
vortices in an ordinary superconductor characterized by 
$\f{|\Psi|^2}{m} = \left(\f{|\Psi_1|^2}{m_1} + \f{|\Psi_2|^2}{m_2}\right)$.
Let us observe that if {\it both} 
 phases $\phi_{1,2}$ change  by $2\pi$ 
around the core then a vortex  carries {\it one}
quantum of magnetic flux.

The vortices characterized by $ \Delta( \phi_1 - \phi_2)   = 4 \pi n$
in the case  where $\f{|\Psi_1|^2}{m_1} \ne \f{|\Psi_2|^2}{m_2}$
have  nontrivial structure.
Let us first consider the case when $\Delta ( \phi_1 +\phi_2)  = 0$
and $ \Delta ( \phi_1 -\phi_2)  = 4 \pi$.
First of all  such a vortex features  a
 neutral superflow characterized by a $4 \pi$ gain in the variable $\gamma$.
In the case of $\eta \ne 0$ this vortex is described by the sine-Gordon functional: 
$F=(1/4) \rho^2 \sin^2\tilde{\theta} (\nabla \gamma)^2 +
\rho^2 K \sin\tilde{\theta} \cos\gamma $.
[In the limit  $\f{|\Psi_1|^2}{m_1} = \f{|\Psi_2|^2}{m_2}$,
this vortex
 does not involve a nontrivial 
configuration of the field $\ccc$, as can be seen from (\ref{new1})].
In the case $K=\eta=0$ this vortex is equivalent to a vortex in a neutral 
superfluid with superfluid stiffness $ (1/2)\rho^2 \sin^2\tilde{\theta}$. 
%(see (\ref{new1})). The most interesting  property of this vortex is 
{ \it As follows  from (\ref{new1}), in the case  when  $\f{|\Psi_1|^2}{m_1} \ne \f{|\Psi_2|^2}{m_2}$,
a topological defect in the neutral field associated
with $\gamma$ is necessarily accompanied by a nontrivial configuration of the 
charged field 
$\ccc$}. % or equivalently besides neutral superflow there 
%exists a nontrivial configuration of a charged supercurrent.
%Let us examine supercurrent  and magnetic field induced by a vortex
%characterized by
%$\Delta ( \phi_1 -\phi_2)  = 0;  \ \  \Delta ( \phi_1 + \phi_2)  = 4 \pi$.
That is, for such a topological defect 
the second  term in (\ref{new1})  becomes
$ \rho^2 \left[  \left\{
 \sin^2\left( \f{\tilde{\theta}}{2}\right) 
- \cos^2\left( \f{\tilde{\theta}}{2}\right) \right\}  
\f{\nabla( \phi_1 - \phi_2)}{2}
- e{\bf A}\right]^2$.
%new2}\eee
This term is nonvanishing 
 for such a vortex configuration, 
which means that 
this  vortex besides  neutral vorticity  also carries
magnetic field.
Lets us calculate the magnetic flux  carried by such a vortex.
The supercurrent around the core of this vortex 
%characterized by  $\Delta ( \phi_1 +\phi_2)  = 0;  \ \  \Delta ( \phi_1 - \phi_2)  = 4 \pi$
is:
$
{\bf J} = 2 e \rho^2 \big[
\big\{ \sin^2\big( \f{\tilde{\theta}}{2}\big) - \cos^2\big( \f{\tilde{\theta}}{2}\big) \big\}  
\f{\nabla( \phi_1 - \phi_2)}{2}
- e{\bf A}\big] 
%\la{newnew}
$.
Lets us now integrate this expression over a closed path $ \sigma $ situated at a
distance  larger than $\lambda$ from the vortex core.
Indeed at a distance much larger than the penetration length the 
supercurrent $\bf J$, or equivalently the massive field $\ccc$, vanishes.
%(We should stress  that this  does not apply to neutral superflow 
%which is also present in the system but which does not 
%vanish exponentially, away from the vortex core.)
Thus we arrive at the following equation:
\beg
\Phi = 
\cos\tilde{\theta} \oint_{\sigma} \f{1}{2 }\nabla( \phi_1 - \phi_2) d{\bf l} =  \cos\tilde{\theta} {\Phi_0} \equiv \tilde{n}_3 \Phi_0,
\la{new3}
\eee
where  $\Phi= \oint_{\sigma} {\bf A} d{\bf l} $ is the magnetic flux 
carried by our vortex and $\Phi_0 = 2\pi  /e$ is  the standard magnetic flux quantum.
{ \it From (\ref{new3})  it follows that in our system such a vortex can carry   arbitrary 
fraction of magnetic flux quantum since it depends on
the free parameter $\cos \tilde{\theta}$
which is a measure of the relative densities of
the  two condensates in the system.  }
In the general case, a vortex characterized by the following phase
changes around the core $\Delta \phi_1 = 2\pi k_1 \ {\rm and} \ \Delta \phi_2 = 2\pi k_2$,
carries the following flux:
\beg
\Phi_{(k_1,k_2)} = 
\Big[\sin^2\Big( \f{\tilde{\theta}}{2}\Big)k_1 + \cos^2\Big( \f{\tilde{\theta}}{2}\Big)k_2
\Big]
 {\Phi_0} 
\la{new31}
\eee

We should emphasis that a straightforward inspection of 
(\ref{e3}) shows that this model allows neutral vortices 
associated with the neutral $O(2)$ boson without a nontrivial 
configuration of the field  $\ccc$ for  any values 
of $\tilde{n}_3$. However
such solutions (e.g. vortices in the case $\tilde{n_3} \ne 0$
characterized
by $\Delta \gamma =2 \pi ; \ccc \equiv 0$) are unphysical 
because  these vortices do not satisfy the
condition that $\phi_i$  change by
$2\pi k_i$ around  the vortex core
(as  follows from (\ref{new1})). 
Thus, while the original model (\ref{act})
and the extended Faddeev model (\ref{e3})
have the same number of degrees of freedom
so that the fields $\bfn$ and $\ccc$
are dynamically independed by construction, 
the mapping incurs a constraint
on topological defects in  $\bfn$ and $\ccc$
since $\phi_{1,2}$ appear in both of them.
%which indeed imposes two conditions:  when we go around the vortex 
%core  the two phases $\phi_i$ should have $2\pi n_i$ gains 
%(with $n_i$ being necessary integer). 
%This 
%should be taken into account in discussion of equivalence mapping 
%between (\ref{act}) and (\ref{e3}) in a non-simply connected 
%space.
So, in the hydrodynamic limit in (\ref{e3})
there is {\it no direct coupling} between the fields $\bfn$
and $\ccc$, 
{\it however}  in a non-simply-connected space a topological
defect in the field $\bfn$  necessary induce 
a nontrivial configuration of the 
field $\ccc$. 
The consequence of this
is the fractionalization of magnetic flux  
in  the model (\ref{e3}).
%As we discussed above this results in the circumstance that 
%in the case when ground state value of the vector $\bfn$
%is not situated on equator of $ S^2$ then the only physical 
%solution for a vortex with neutral superflow necessarily involves
%nontrivial configuration of  the field $\ccc$ and necessarily 
%carries a fraction of magnetic flux quantum.}
%Let us discuss details of the structure of the  vortex 
%characterized by $\Delta ( \phi_1 -\phi_2)  = 0;  \ \  \Delta ( \phi_1 + \phi_2)  = 4 \pi$.
% For simplicity  we can  consider the case when
% $\xi_1 \approx \xi_2 << \lambda$.
%The  magnetic field  in this vortex at $r > \xi$ behaves as:
%\be
%H=  \f{\cos\theta \Phi_0}{2\pi \lambda^2} K_0\left(\f{r}{\lambda}\right)
%\la{mf}
%\ee
%Where $K_0$ is the Macdonald function, and $\Phi_0=\pi\hbar c/e$ is the magnetic flux quantum.

So a two-band superconductor (\ref{act})
and extended Faddeev model (\ref{e3}) allow the following linear 
composite topological defects:
%\begin{enumerate}

\underline{(i) $\Delta (\phi_1-\phi_2)  = 4\pi n$; \ $\Delta(\phi_1+\phi_2) = 0$}:  these are the 
vortices which 
 feature  neutral superflow. When the ground state of the 
variable $\bfn$ corresponds to the equator on $S^2$  (that is 
the case when $\f{|\Psi_1|^2}{m_1}=\f{|\Psi_2|^2}{m_2} $ or
equivalently  $\cos \tilde{ \theta} = 0 $)
these  vortices do not carry  magnetic flux. In the case 
when $\cos \tilde{ \theta}  \ne 0$, 
%the non-simply-connectedness  of space in the presence of linear defects 
%results in the fact that 
the only physical 
solutions with neutral vorticity
% also involve nontrivial configuration of the field $\ccc$  and such vortices
also  carry a fraction of magnetic flux quantum. % $\Phi=\cos \tilde{\theta} \Phi_0$. 
  In the case of nonzero Josephson coupling these vortices
are described by the sine-Gordon equation (\ref{new1}).% When
%$\eta \rightarrow 0$  these vortices resemble vortices in a neutral superfluid.

\underline{(ii) $\Delta \gamma = 0 $; \ $\Delta(\phi_1+\phi_2) = 4\pi k$}
these are the vortices which 
feature circular supercurrent and no circular neutral superflow. Such vortices 
carry $ k $ flux quanta  of magnetic field. 
%{ \it When Josephson
%coupling is negligible small these are the only type 
%linear vortices with finite energy per unit length in the systems (\ref{act})
%and (\ref{e3}).} 
%From our analyses it 
%follows that only this  type of vortices can be induced in the system 
%by applied external magnetic field \cite{com6}. 
{ \it One of the  
physical consequences of our analysis is that we can
observe that  in  the
two-gap superconductor  (\ref{act}), (\ref{e3}), 
%with small Josephson coupling %it is not allowed
in spite of the existence of two types of Cooper pairs,
a formation of two sublattices of vortices corresponding
to each of the condensates
in an external field is  
 energetically forbidden. This is because the
energy per unit length  of noncomposite vortices 
is divergent in an infinite sample both in cases of zero and nonzero Josephson
coupling (in  case of finite $\eta$ 
a vortex creates a domain wall
 which makes
its energy per unit length divergent 
in infinite sample \cite{newa,com6}).}
%That is, in case of large Josephson coupling it is forbiden by ``phase-locking"
%term $K \sin\tilde{\theta} \cos \gamma$ and 
%in the limit of  negligible small $K$ is is forbidden by the fact that as
%follows from (\ref{new1}) only composite vortices 
%of type  (ii)
% have  finite energy per unit length.
We can also  observe  that   a superconductor 
made up of  one condensate which is of type-II and another 
condensate  which is of type-I will
in general, in the presense of an
external magnetic field,   
form one flux quantum vortices involving both condensates 
and will preserve two-gap  superconductivity, 
even if the external field exceeds the thermodynamic critical 
magnetic field for the type-I condensate. 

\underline{(iii) $\Delta \gamma = 2\pi n$; \ $\Delta(\phi_1+\phi_2) = 4 \pi l $}:
These are the vortices which   may be viewed
as a co-centered $l$-flux quanta  Abrikosov vortex
combined with  a vortex with neutral vorticity with winding number 
$n$ carrying  a fractional magnetic flux. These vortices
and the vortices of type (ii) characterized by 
$\Delta(\phi_1+\phi_2)  \geq  8 \pi  $ 
could be unstable against decay into more simple vortices
\cite{newa}.

Besides composite vortices  the system allows vortices
where the phase of one of the  condensates changes by $2\pi$ 
around the core whereas the phase of the second condensate remains
constant. {\it These vortices  are also principally different from ordinary 
Abrikosov vortices}. Let us discuss these vortices in the 
Ginzburg-Landau formalism (\ref{act}). 
This will  provide additional perspective on the discussion 
and will allow us to illustrate the connection between the models (\ref{act})
and (\ref{e3}). We shall present a solution for these vortices in the limit 
when the inverse mass for $n_1$ which is given by Josephson term 
is larger than the sample size. % and smaller than magnetic field penetration length.
   Such an approximation allows one to
put $\eta$ to zero.  
%We should emphasize that the strength of interband Josephson
%coupling in two-band superconductors such as $MgB_2$
%has not yet been established in experiments and 
%it is extremely difficult to measure \cite{legg}. 
%The described below
%the nature of vortex excitations  indeed could
%directly answer the question if the approximation 
%when mass for neutral boson is smaller than inverse
%penetration length holds true in  e.g. $Mg B_2$. 
%Our observation is that the nature of topological excitations
%in a two-band material depends is directly coupled to the strength intrinsic Josephson effect
%and it could serve as a direct possibility
%to deduce from experiments the strength of 
%phase-locking effects in two-band materials. }
We should note  that there certainly exist systems where 
$\eta$ is exactly zero. That is, a model with $\eta=0$ should 
describe vortices in a bi-layer system:  
a superconductor-insulator-superconductor 
compound
is an example of a system of two condensates
coupled only by a gauge field. In such a system, a
vortex in one layer carries flux through the second layer,
where it also induces a current,
which should lead   to the flux fractionalization.
Besides that an additional $U(1)$ symmetry appears if one considers 
equal-spin pairing.  In that case, if the spin-orbit coupling is neglected, each
spin population has its own phase, without Josephson coupling \cite{vv}.

Let us recall the GL equation  for the supercurrent in standard notations:
$
{\bf J} = \f{i e}{2m_1}[\Psi_1^*\nabla \Psi_1 - \Psi_1\nabla\Psi_1^*]
+ \f{i e}{2m_2}[\Psi_2^*\nabla \Psi_2 - \Psi_2\nabla\Psi_2^*]
+e^2 {\bf A}\left[ \f{|\Psi_1|^2}{m_1}+\f{|\Psi_2|^2}{m_2}\right]
$. In  the situation 
when only the phase $\phi_1$ changes by $2 \pi$ 
around the core while the phase of the second
condensate remains constant, we have the following expression
for the vector potential:
$
{\bf A} ={\bf J} \left[ e^2 \left(  \f{|\Psi_1|^2}{m_1}
+ \f{|\Psi_2|^2}{m_2} \right) \right]^{-1}+ 
\f{1}{e}{\qq}\left[{\qq+ \ww}\right]^{-1} \nabla\phi_1
$.
From this expression it follows that 
the vortex characterized by $\Delta \phi_1 = 2 \pi; \Delta \phi_2 =0 $
carries the following fractional magnetic flux
\beg
\Phi=\oint {\bfA} d l={\qq} \left[{\qq+ \ww}\right]^{-1} \Phi_0,
\eee
%where $\Phi_0=\f{2\pi}{e}$ is standard flux quantum.
%Thus in this fundamental property the two-gap
%superconductor is principally different from 
%the ordinary one-gap system \cite{frac}.
Let us now  remark on what are the physical roots of  this flux 
fractionalization.
Further examining the solution for 
the vortex ($\Delta \phi_1=2\pi, \Delta \phi_2=0$),
we can write the vector potential 
as  ${\bf A} = \f{{\bf r}\times {\bf e}_z}{|r|}|{\bf A}(r)|$
where $r$ measures the distance from the core 
and ${\bf e}_z$ is a unit vector pointing along the core.
The magnetic field is then given by
$ |{\bf B}|=\f{1}{r} \f{d}{d r}(r |{\bf A}|)$.
The equation for the current $\bf J$ can then be rewritten
as 
\beg
-\f{d}{dr}\Bigl[\f{1}{r}\f{d}{d r}(r|{\bf A}|) \Bigr]
+\qq\Bigl[|{\bf A}|e^2-\f{e}{r}\Bigr] +\ww|{\bf A}|e^2 =0
\nonumber
\eee
For such a vortex, the solution for the vector potential is 
\beg
|{\bf A}|&=&\f{\qq}{\qq+\ww}\f{1}{er}- \nonumber \\
&&\f{\qq}{\sqrt{\qq+\ww}}K_1\Bigg(e \sqrt{\qq+\ww} r\Bigg)
\label{A}
\eee
Indeed the 
magnetic field vanishes exponentially away 
from the vortex core
with the characteristic length scale given by the magnetic
field penetration length $\lambda=\Big[e \sqrt{\qq+\ww}  \Big]^{-1}$: 
$|{\bf B}|=e\qq K_0\left(e \sqrt{\qq+\ww} r\right)$.
Also, in contrast to the Abrikosov vortex \cite{aaa},
besides having the fractionalization  of magnetic flux 
our vortex also features the 
neutral vorticity. This, in particular, 
%in GL formalism 
can be
seen by substituting the solution  (\ref{A})
into (\ref{act}). Then, at  length scales larger than the
magnetic field penetration length from 
the vortex core, we have 
the following expression for the energy density:
$
F= \f{1}{2m_1}\bigg|\bigg( \nabla + i{\qq}\bigg[{{\qq+\ww}}
\bigg]^{-1/2}\f{1}{r}\bigg)\Psi_1 \bigg|^2
%\nonumber \\
+
\f{1}{2m_2}\bigg|\bigg(  i{\qq}\bigg[{{\qq+\ww}}
\bigg]^{-1/2}\f{1}{r}\bigg)\Psi_2 \bigg|^2$.
%\label{nnn} \eee
Thus the energy  per unit length of the
 vortex ($\Delta \phi_1=2\pi, \Delta \phi_2=0$)
 is divergent. This is due to the fact that 
such a topological configuration
necessarily induces in a two-gap system 
the  neutral superflow.
Indeed the above expression for $F$
is similar to the energy density 
of a vortex 
in a {\it neutral} system 
%with a vortex  in a {\it neutral} phase field $\phi_1$
with effective stiffness $\qq \ww \Big[ \qq+\ww\Big]^{-1}$:
\be
F_n= \f{1}{2}
{\qq \ww}\Bigg[{\qq+\ww}\Bigg]^{-1}
| \nabla \e^{i \phi_1} |^2
\label{lll}
\ee
%Let us remark that the procedure which 
%we arrive at Eq. (\ref{lll}) is equivalent 
%to that used in the first part of the paper.
%with the only difference
%that in the first part of the paper,  first the variables
%were separated  into a neutral and a charged fields
%(the London limit of the general procedure \cite{we}), 
%and then there were found general solutions
%for vortices $S^1 \rightarrow S^1$
%while in the second part of this  paper  first
%the solution found for a vortex 
%($\Delta \phi_1=2\pi, \Delta \phi_2=0$) was
%substituted in (\ref{act})
%and then the divergent part associated
%with neutral vorticity extracted from both kinetic terms 
%in (\ref{act}).
This shows transparently the physical origin 
of the presence in the system of a massless
neutral boson; {\it   a topologically nontrivial
configuration  ($\Delta \phi_1=2\pi, \Delta \phi_2=0$),
besides having a current in the condensate $\Psi_1$ also necessarily
induces  current in the condensate $\Psi_2$.
Albeit in such a configuration there are no gradients of $\phi_2$,
however the two condensates are not independent
but are connected by the vector potential.} 
The admixture of an oppositely directed supercurrent of the 
condensate $\Psi_2$  
%which necessarily accompanies such a vortex in two-gapsystem, 
leads to the situation when the  two supercurrtents
partially compensate the induced by each other magnetic 
field. This  leads
to  the  existence of the effective neutral superflow
in the system (which in particular should dominate
long-range interaction between vortices). Moreover it is the fact that
the two currents 
partially compensate induced by each other magnetic
field, which leads to the %basic property of the two-component Abelian Higgs model:
fractionalization of magnetic flux.
We  stress that the vortex solutions
  ($\Delta \phi_1=2\pi, \Delta \phi_2=0$) in
this model, albeit being topologically
stable, however can not form as an energetically
preferred state in external field. In an external field
the system will form the composite
vortices   ($\Delta \phi_1=2\pi, \Delta \phi_2=2\pi$) described above.

%It is a remarkable fact that addition 
%of a new component in the Abelian
%Higgs model  deeply changes the basic properties of the model.
There is however an experimental set up 
which should allow one to directly observe ($\Delta \phi_1=2\pi, \Delta \phi_2=0$) 
 excitations.
That is, in a two - gap superconducting film described by 
(\ref{act})  there should occur thermal creation of 
pairs of vortices and antivortices
  with neutral vorticity and fractional magnetic flux.
%In the case if the ground state value $\tilde{n}_3$
%is not located too close to north or south poles of $S^2$
%the  interaction between these vortices will be 
%dominated by a neutral superflow. 
The fact that this type of vortex 
features  both  confined magnetic field  and  long-range interaction due
to  neutral superflow will allow: (i) detection of the   Berezinskii-Kosterlitz-Thouless
 transition in a system 
of these vortices by standard experimental techniques like flux-noise measurements 
or measurements in an applied  dynamic magnetic field (like in 
a system of Abrikosov vortices  \cite{minnhagen}).
(ii) such measurements 
 straightforwardly allow one to extract data about type of  interaction between 
a vortex and an  antivortex in the system. { \it In spite of being
 by definition sensitive only to magnetic flux 
carrying vortices  these measurements should 
give exactly the picture of the BKT transition like in 
a neutral $U(1)$ system}. That is, the transition should be  a pure 
BKT transition but not a ``would be" BKT transition 
like in a system of Abrikosov vortices. As it is well known there is no 
true BKT transition is a system of Abrikosov vortices:
because  
their interaction is screened by the Meissner effect 
 and a BKT transition is replaced
by a crossover which is observable only when penetration length is
sufficiently  large \cite{pearl}.   
{ \it Moreover the BKT transition  in a system of these vortices should be observable even
%in flux noise and applied dynamic magnetic field measurements 
%in spite the interaction between these vortices 
%is dominated by  neutral superflow and even 
 in a type-I system }
(both in the limits $\eta =0$
and when   $\eta$ is large, where one has 
sine-Gordon vortices interacting with a linear potential \cite{newa}).
%Also such an experiment 
%involving interaction of topological defects
%is a direct probe to the difficult question 
%of measurement of the 
%importance of  ``interband phase-locking effects" 
%in a two-band system.
% More details will be presented elsewhere \cite{newa}.
%The only necessary condition 
%for observation of this phenomenon is $0<|\tilde{n}_3|<1$
%and smallness of phase-locking term (see also comment \cite{comme})
%.

In conclusion, we studied vortices 
in a two-gap superconductor. 
%We identified the most energetically preferred
%solution which should form in the system in applied magnetic field.
%Besides that 
We have shown that the system allows 
several  types of vortices,  such as 
vortices carrying an arbitrary fraction of the magnetic flux 
quantum, which have no counterpart in ordinary
one-gap superconductors or two-gap neutral  superfluids. 
%We proposed an experimental set up which may confirm existence of
%such vortices in  $Mg B_2$,
%or transition metals  $Nb$, $Ta$, $V$ and $Nb$-doped $Sr Ti O_3$.
% two-band superconductor (which is exactly equivalent to  the
%double-condensate Ginzburg-Landau model \cite{we}). 
%In the first part we discussed linear topological defects allowed by this system
%and have shown that they are principally different 
%from Abrikosov vortices in one-gap superconductor 
%due to coexistence  in our system of a massless neutral
%boson and a Meissner effect. 
%Our analysis shows  again (along with the  recent results \cite{we}  
%on duality between two-gap Ginzburg-Landau model and 
%extended Faddeev model and observation of the fact that 
%the system allows formation of knot solitons)   that the two-gap superconductivity (or equivalently 
%two-flavour Abelian Higgs model) in spite 
%of its simplicity on the first glance, is actually  a nontrivial model 
%with numerous hidden phenomena which do not parallel one-gap superconductivity.
This study  also has interdisciplinary interest because 
similar models are highly relevant  in particle physics  \cite{tdlee,nature}.

It is a great pleasure to thank  L.D. Faddeev, D. Gorokhov, A. J. Niemi, 
G. Volovik, K. Zarembo, V. Cheianov, Y.M. Cho, O.Festin, S. Girvin and P. Svedlindh 
for  interesting discussions and/or comments. 
This  work  has
been supported by grant STINT IG2001-062, the Swedish
Royal Academy of Science and Goran Gustafsson Stiftelse UU/KTH.
The author also acknowledges NorFA mobility scholarship and 
Thordur Jonsson and Larus Thorlacius for hospitality 
during his stay at University of Iceland.
 
%\begin{figure}[p]
%\hspace{0cm}\epsfxsize=400pt\epsfbox{theta.eps}
\end{multicols}

\begin{thebibliography}{99}
\bibitem{com6}
The energy of the fractional flux  vortices 
in a such a  system will however be finite in a mesoscopic sample.
Also, in a granular superconductor, lattice effects may rise
mass of the neutral $O(2)$ boson  which will make 
the energy per unit length of this defect finite.
\end{thebibliography}
\end{document}